# Possibility of Emergence of Chiral Magnetic Soliton in Hexagonal Metal Formate $[NH_4][M(HCOO)_3]$ with $M^{2+}$ = Mn, Fe, Co, and Ni and $KCo(HCOO)_3$

**L. M. Volkova[1] and D. V. Marinin[1]**

**Abstract** Search of chiral magnetic solitons on the basis of the data on the crystal structure is important from both fundamental and technological points of view. Potential chiral magnetic solitons - hexagonal metal formates $[NH_4][M(HCOO)_3]$ with $M^{2+}$ = Mn, Fe, Co, and Ni and $KCo(HCOO)_3$ containing dominating left-handed chiral antiferromagnetic helices that compete with interhelix interactions have been identified. The sign and strength of magnetic interactions not only between nearest neighbors, but also for longer-range neighbors in the metal formates have been calculated on the basis of structural data. Magnetic structures of these compounds are similar to that of the chiral magnetic soliton $Cr_{1/3}NbS_2$. The interrelation between chiral polarization of crystalline and magnetic structures in the metastable polymorph $KCo(HCOO)_3$ has been demonstrated.

## 1 Introduction

Both chirality [1, 2] and solitons [3-5] phenomena have very important roles as in nature as in research. When united in the field of magnetism in the form of chiral magnetic solitons, they become rather promising materials for spintronics. Spin solitons are solitary waves of the spin density [5, 6]. Chirality (from the Greek word “ceir” - “hand”) is the object property to be incompatible with its mirror reflection by any combination of rotations and displacements in a three-dimensional space. In magnetism, chirality means the left- or right-handedness associated with the helical (spiral) order of magnetic moments. The factors controlling the emergence of chiral magnetic solitons include Dzyaloshinkii-Moriya (DM) antisymmetric exchange interaction, spin frustration, and magnetic anisotropy [5-16].

As we showed in [17], the dominating antiferromagnetic (AFM) magnetic couplings in $Cr_{1/3}$NbS2 emerged on the left-handed crystallographic spiral chain of magnetic $Cr^{3+}$ atoms. The directions of magnetic moments (spins) of adjacent atoms in these left-handed spiral chains are non-collinear but slightly canted because of the chain spiral geometry. Non-collinearity of spins in chiral chains, along with the competition of strong intrachain AFM interactions with weaker interchain AFM ones, induce the emergence of a long-period helical spin ordering in chiral chains. In other words, chiral spin helixes are thus created. The role of DM interaction в $Cr_{1/3}NbS_2$ consists of final ordering and stabilization of chiral spin helices into a chiral magnetic soliton lattice, i.e., into a chiral magnetic soliton (chiral helimagnet).

Determination of crystal chemistry conditions for the emergence of the chiral magnetic soliton lattice and search of compounds – potential chiral magnetic solitons – for which these conditions are fulfilled, in the Inorganic Crystal Structure Database (ICSD, FIZ Karlsruhe, Germany) is very important from both fundamental and technological points of view.

It is generally accepted that in noncentrosymmetric crystals the chiral heliomagnetic structure is formed by relativistic spin-orbital DM interaction [5-16] through selection of one of two helices (left- or right-handed). In the absence of DM interaction, the magnetic structure inevitably becomes achiral. So it turns out that the crystal structure has a modest role in this process, namely elimination of the center of symmetry, because the DM interaction manifests itself only in noncentrosymmetric structures. However, there exist many

[1] Institute of Chemistry, Far Eastern Branch of Russian Academy of Science, 159, 100-Let Prosp., Vladivostok 690022, Russia
e-mail: volkova@ich.dvo.ru

non-centrosymmetric magnetic compound whereas chiral magnetic solitons are rather scarce. Therefore, the mentioned condition is necessary, but not sufficient for the chiral soliton formation

For example, it was established that in the emergence of the chiral soliton in $CsCuCl_3$ [5, 18-20] the important role, aside from the DM interaction, belonged to helical chiral chains of magnetic ions of divalent copper originated from the Jan-Teller effect. At the same time, the emergence of the chiral magnetic soliton in the two-dimensional intercalate $Cr_{1/3}NbS_2$ [21], in which magnetic ions of trivalent chromium form flat triangular lattices, is usually attributed to the relativistic effect of Dzyaloshinskii-Moriya. Possibly, the latter was the reason of the formed opinion about the existence of structurally and non-structurally originated solitons.

On the other hand, from the practical point, the chiral magnetic soliton $Cr_{1/3}NbS_2$ is the most interesting, since, as was found, it contained self-organized spin helices whose twist degree is controlled by the magnetic field [8, 22]. The latter opens exciting possibilities of using the molecular magnetism in creation of magnetically controlled chiral nano-objects and devising the principally new three-dimensional memory devices on their basis [23-25]. That is why we selected the chiral magnetic soliton $Cr_{1/3}NbS_2$ as a prototype for prediction and determined the role of structural factors in formation of the chiral magnetic soliton lattice [17].

Based on the determination of crystal chemistry solutions of the emergence of chiral magnetic soliton lattice in $Cr_{1/3}NbS_2$ [17] and $CsCuCl_3$ [26], we formulated crystal chemistry criteria for search of potential solitons, according to which their crystal structures must have, aside from the necessary absence of the symmetry center, two more important characteristics: first, the presence of chiral spin helices as the base of the chiral magnetic soliton lattice and second, these chiral spin helices must be dominating in the system and their competition with other interactions must promote the formation of superstructures with large periods. The formation of chiral spin helices is caused by the compound crystal structure in combination with magnetism.

Search of potential chiral magnetic solitons on the basis of structural data is crucial to perform not only among the compounds having a quasi-one-dimensional crystal structure, since noncoincidence of crystal and magnetic structures comprises a rather widely spread phenomenon. The latter is related to the fact that the sign and strength of exchange magnetic interactions and, therefore, the dimensionality of magnetic structures are mainly determined by the size of magnetic and intermediate ions and the geometry of location of intermediate ions in the local space between magnetic ones. Here, a good example is the presence of a quasi-one-dimensional magnetic structure formed of chiral helices stretched along the *c* axis as in layered $Cr_{1/3}NbS_2$ as in chain $CsCuCl_3$ compounds.

The objective of the present work was to find in the ICSD database potential chiral magnetic solitons similar to those of $Cr_{1/3}NbS_2$ and suggest them for the experimental study of the spin structure in a magnetic field.

## 2 Materials and Methods

### 2.1 The Method of Calculation

The search of potential chiral magnetic solitons was performed in the ICSD database among magnetic compounds crystallizing in the same non-centrosymmetric hexagonal space group $P6_322$ as $Cr_{1/3}NbS_2$. To determine the characteristics of magnetic interactions, we used the earlier developed [27, 28] phenomenological crystal chemistry method and the '*MagInter*' program created on its basis.

The method enables one to determine the sign (type) and strength of magnetic couplings on the basis of structural data. According to this method, a coupling between magnetic ions $M_i$ and $M_j$ emerges in the moment of crossing the boundary between them by an intermediate $A_n$ ion with the overlapping value of ~0.1 Å. The area of the local space between the ions $M_i$ and $M_j$ along the bond line is defined as a cylinder, whose radius is equal to these ions radii. The strength of magnetic couplings and the type of magnetic moments ordering in insulators is determined mainly by the geometrical position and the size of intermediate $A_n$ ions in the local space between two magnetic ions $M_i$ and $M_j$. The positions of intermediate $A_n$ ions in the local space are determined by the distance $h(A_n)$ from the center of the ion $A_n$ up to the bond line $M_i$-$M_j$ and the degree of the ion displacement to one of the magnetic ions expressed as a ratio ($l_n'/l_n$) of the lengths $l_n$ and $l_n'$ ($l_n \le l_n'$; $l_n' = d(M_i\text{-}M_j)\text{-}l_n$) produced by the bond line $M_i$-$M_j$ division by a perpendicular made from the ion center.

The intermediate $A_n$ ions will tend to orient magnetic moments of $M_i$ and $M_j$ ions and make their contributions $J_n$ into the emergence of AFM or ferromagnetic (FM) components of the magnetic interaction in dependence on the degree of overlapping of the local space between magnetic ions ($\Delta h(A_n)$), the asymmetry ($l_n'/l_n$) of position relatively to the middle of the $M_i$-$M_j$ bond line, and the distance between magnetic ions ($M_i$-$M_j$).

Among the above parameters, only the degree of space overlapping between the magnetic ions $M_i$ and $M_j$ ($\Delta h(A_n) = h(A_n) - r_A$) is equal to the difference between the distance $h(A_n)$ from the center of $A_n$ ion up to the bond line $M_i$-$M_j$, and the radius ($r_A$) of the $A_n$ ion determined the sign of magnetic interaction. If $\Delta h(A_n)<0$, the $A_n$ ion overlaps (by $|\Delta h|$) the bond line $M_i$-$M_j$ and initiates the the emerging contribution into the AFM

**Table 1** An estimate of $J_n$ magnetic couplings in oxides $Mn^{2+}$, $Fe^{2+}$, $Co^{2+}$ and $Ni^{2+}$ by crystal chemical method (I) and experimental and quantum-chemical methods (II)

| Compound | Space group, lattice parameters | d(M-M) (Å) | $J$ (Å$^{-1}$) I (This work) | $J$ (meV) II | $K^a$ | $K^a \times J$(Å$^{-1}$) (meV) |
|---|---|---|---|---|---|---|
| MnWO4 [29] ICSD-67906 | *P*2/*c* (No. 13) *a* = 4.830, *b* = 5.760, *c* = 4.994 Å 90. 91.14. 90 | 3.286 4.830 | 0.0356 (FM) 0.0324 (FM) | 0.42 (FM) [30] 0.32 (FM) [30] | 11 | 0.39 (FM) 0.36 (FM) |
| $SrFeO_2$ [31] ICSD-418606 | *P*4/*mmm* (N123) *a* = 3.991, *b* = 3.991, *c* = 3.475 Å | 3.991 | 0.1758 (AFM) | 7.04 (AFM) [31,32] | 40 | 7.03 (AFM) |
| $Co_3V_2O_8$ [33-35] ICSD-2645 | *Cmca* (No. 64) *a* = 6.030, *b* = 11.486, *c* = 8.312 Å 90. 90. 90 | 2.987 | 0.0385 (FM) | 1.25 (FM) [35] | 32.5 | 1.25 (FM) |
| $Ni_3TeO_6$ [36] ICSD-240377 | *R*3*H*(No. 146) *a* = 5.109 *b* = 5.109 *c* = 13.767 Å 90. 90. 120 | 3.003 | 0.0233 (FM) | 0.94 (FM) [37] | 40.34 | 0.94 (FM) |

[a]Scaling factor for translating the value per angstrom$^{-1}$ into meV in oxides $M^{2+}$

emerging contribution into the AFM component of magnetic interaction. If $\Delta h(A_n)>0$, there remains a gap (the gap width $\Delta h$) between the bond line and the $A_n$ ion, and this ion initiates a contribution to the FM component of magnetic interaction.

The sign and strength of the magnetic coupling $J_{ij}$ is determined by the sum of the above contributions:

$$J_{ij} = \sum_n j_n \, .$$

The value $J_{ij}$ is expressed in units of Å$^{-1}$. If $J_{ij} < 0$, the type of $M_i$ and $M_j$ ions magnetic ordering is AFM and, in opposite, if $J_{ij} > 0$, the ordering type is FM.

The comparison of our data with that of other methods shows that the scaling factors for translating value angstrom$^{-1}$ into meV in oxides $Mn^{2+}$, $Fe^{2+}$, $Co^{2+}$ and $Ni^{2}$ are 11, 40, 32.5 and 40.34, respectively (Table 1).

### 2.2 Search of Chiral Magnetic Solitons

We calculated the sign and strength of magnetic interactions not only between nearest neighbors but also for longer-range neighbors in 20 magnetic compounds selected from the crystal chemistry point. These compounds had different compositions and structural types. As the initial data for calculations, we used crystallographic parameters and atomic coordinates of crystal structures stored in *cif* files and ionic radii of atoms determined in [38], except the radius of the carbon atom in $sp^2$ hybridization in formate ligands ($HCOO^-$), for which the covalent radius (r = 0.73 Å) was taken from [39].

According to our calculations, only 5 among the examined compounds contain dominating chiral antiferromagnetic (AFM) helices and magnetic structures of these compounds are similar to that of $Cr_{1/3}NbS_2$. The above compounds are metal formates $[NH_4][M(HCOO)_3]$ with $M^{2+}$ = Mn, Fe, Co, Ni [40, 41] and $KCo(HCOO)_3$ [42] assigned to two structural types: $NH_4Mn(HCOO)_3$ and $KCo(HCOO)_3$, respectively.

On the basis of structural data, we calculated the sign and strength of magnetic couplings ($J_{ii}$) in 14 samples of the metal formate frameworks of $[NH_4][M(HCOO)_3]$ with $M^{2+}$ = Mn, Fe, Co, Ni, and $KCo(HCOO)_3$ (data for ICSD 181920-181922 [42]; 240672-240674 [40]; 262006-262012 [41]). Calculations were performed for 2 phases ($P6_322$ and $P6_3$) at 290(293)K and 110K, respectively. Besides, characteristics of magnetic couplings in the $C2/c$ phase of $KCo(HCOO)_3$ (data for ICSD 181923 [42] were calculated**.**

Table 2 shows the results of these calculations only for eight samples from [41, 42] (the remaining data are about the same as those shown) and the respective parameters of the soliton $Cr_{1/3}NbS_2$ (at low temperatures) from [17] for comparison.

In the $P6_322$ phase, the multiplicity of magnetic couplings $J1_1$, $J1_2$, $J2$, $J3$, and $J4$ is equal to 6; for $J5$ and $J6$ couplings – to 12; and for $J_c$ couplings – to 2. At the transition from $P6_322$ to $P6_3$, each of the groups of $J1_1$–$J4$ couplings is split according to symmetry elements into three and, in case of $J5$ and $J6$ couplings, into six subgroups with multiplicity equal to 2. In all these groups/subgroups (except $J6$), magnetic couplings are crystallographic and magnetic equivalents, while each of $J6$ groups (subgroups) is additionally split, despite symmetry elements, into two groups of $J6$ and $J6'$ couplings (their magnetic coupling strengths are substantially different), albeit remaining crystallographically equivalent. Strong AFM $J6$ couplings form left-handed helices, whereas weak $J6'$ couplings form right-handed helices along the $c$ axis. Table 2 does not indicate characteristics of each individual coupling for the $P6_3$ phase, but provides their limits in groups, into which $J1$–$J6$ couplings are split at transition from $P6_322$ to $P6_3$ phase.

Note that our calculations (Table 2) do not show significant differences in parameters of magnetic interactions $J_{ij}$ at the structural phase transition from $P6_322$ to $P6_3$.

**Table 2** Crystallographic characteristics and parameters of magnetic couplings (*Jn*) calculated on the basis of structural data and respective distances between magnetic ions in *P*$6_3$22 and *P*$6_3$ phases of [$NH_4$][M$(HCOO)_3$] ($M^{2+}$ = Mn, Fe, Co, Ni), K$Co(HCOO)_3$, and $Cr_{1/3}NbS_2$ (for comparison)

| | Compound | | | | | | | | | |
|---|---|---|---|---|---|---|---|---|---|---|
| | $Cr_{1/3}NbS_2$ | [$NH_4$][Ni$(HCOO)_3$] | [$NH_4$][Co$(HCOO)_3$] | | [$NH_4$][Fe$(HCOO)_3$] | | [$NH_4$][Mn$(HCOO)_3$] | | K$Co(HCOO)_3$ | |
| References | [25, 47] | [40] | [40] | | [40] | | [40] | | [41] | |
| Data for ICSD | 626392 | 262012 | 262010 | 262011 | 262008 | 262009 | 262006[a] | 262007[a] | 181920 | 181923 |
| T (K) | low T | 293K | 290K | 110K | 290K | 110 K | 290K | 110K | 110K | 293K |
| Space group | *P*$6_3$22 | *P*$6_3$22 | *P*$6_3$22 | *P*$6_3$ | *P*$6_3$22 | *P*$6_3$ | *P*$6_3$22 | *P*$6_3$ | *P*$6_3$22 | *C*2/*c* |
| *a* (Å) | 5.741 | 7.2861 | 7.3058 | 12.5871 | 7.3236 | 12.6167 | 7.3622 | 12.6685 | 6.9978 | 10.7489 |
| *b* (Å) | 5.741 | 7.2861 | 7.3058 | 12.5871 | 7.3236 | 12.6167 | 7.3622 | 12.6685 | 6.9978 | 8.9864 |
| *c* (Å) | 12.097 | 8.0207 | 8.1897 | 8.2237 | 8.3180 | 8.3647 | 8.4885 | 8.5374 | 8.4687 | 6.8879 |
| α β γ (º) | 90 90 120 | 90 90 120 | 90 90 120 | 90 90 120 | 90 90 120 | 90 90 120 | 90 90 120 | 90 90 120 | 90 90 120 | 90 95.47 90 |
| Method[(a)], | XDS | XDS | XDS | XDS | XDS | XDS | XDS | XDS | XDS | XDS |
| R-value[(b)] | | 0.0168 | 0.0208 | 0.0297 | 0.0232 | 0.0287 | 0.0202 | 0.0317 | 0.022 | 0.0267 |
| M | $Cr^{+3}$ | $Ni^{+2}$ | $Co^{+2}$ | $Co^{+2}$ | $Fe^{+2}$ | $Fe^{+2}$ | $Mn^{+2}$ | $Mn^{+2}$ | $Co^{+2}$ | $Co^{+2}$ |
| *r* (Å) | 0.615 | 0.69 | 0.745 | 0.745 | 0.78 | 0.78 | 0.83 | 0.83 | 0.745 | 0.745 |
| **1NN** | | | | | | | | | | |
| d(M-M) (Å) | 5.741 | 7.2861 | 7.3058 | 7.209 – 7.326 | 7.3236 | 7.210 – 7.359 | 7.3622 | 7.230 – 7.399 | 6.998 | 7.005, 8.986 |
| $J1_1^1$ - $J1_1^3$ [(c)] (Å$^{-1}$) | -0.0041 | -0.0002 | 0.0002 | 0.0002 – 0004 | 0.0008 | 0.0003 – 0.0009 | 0.0016 | 0.0014 – 0.0018 | 0.0002 | -0.0012, 0.0297 |
| **$1_2$NN** | | | | | | | | | | |
| d(M-M) (Å) | 11.482 | 14.572 | 14.612 | 14.476 – 14.593 | 14.647 | 14.494 – 14.643 | 14.725 | 14.544 – 14.712 | 13.995 | 14.011 |
| $J1_2^1$ - $J1_2^3$ (Å$^{-1}$) | -0.0032 | 0.0069 | 0.0085 | 0.0084 – 0.0092 | 0.0101 | 0.0098 – 0.0108 | 0.0117 | 0.0071 – 0.0129 | 0.0092 | 0.0061 – 0.0169 |
| **2NN** | | | | | | | | | | |
| d(M-M) (Å) | 6.897 | 5.812 | 5.878 | 5.875 | 5.931 | 5.931x6 | 6.007 | 6.005 | 5.853 | 5.661, 6.101 |
| $J2^1$ - $J2^3$ (Å$^{-1}$) | -0.0077 | -0.0245 | -0.0214 | -0.0203 to -0.0218 | -0.0184 | -0.0162 to -0.0186 | -0.0142 | -0.0111 to -0.0139 | -0.0201 | 0.0538, 0.0581 |
| *j*(C1) - *j*(C3)[(d)] (Å$^{-1}$) | | -0.0185 | -0.0164 | -0.0154 to -0.0170 | -0.0146 | -0.0124 to -0.0147 | -0.0118 | -0.0087 to -0.0115 | -0.0161 | -0.0814, -0.0383 |
| MC1M[(e)]-MC3M[(e)] | | 163.63º | 162.74° | 162.07º – 163.07º | 161.89 º | 160.39º – 161.97º | 160.46º | 158.41º – 160.23º | 162.32º | 169º(O), 179º(C) |
| **3NN** | | | | | | | | | | |
| d(M-M) (Å) | 8.974 | 9.320 | 9.377 | 9.345 | 9.424 | 9.394 | 9.502 | 9.464 | 9.122 | |
| $J3^1$ - $J3^3$ (Å$^{-1}$) | 0.0168 | -0.0022 | -0.0022 | -0.0023 | -0.0022 | -0.0023 to 0.0029 | -0.0022 | -0.0029 | -0.0042 | |
| **4NN** | | | | | | | | | | |
| d(M-M) (Å) | 9.944 | 12.620 | 12.654 | 12.587 | 12.685 | 12.617 | 12.752 | 12.669 | 12.121 | 10.749 – 14.512 |
| $J4^1$ - $J4^3$ (Å$^{-1}$) | -0.0056 | 0.0002 | 0.0002 | 0.0002 | 0.0002 | 0.0003 | 0.0004 | 0.0004 – 0.0005 | 0.0002 | 0.0066 – 0.0180 |

| | Compound | | | | | | | | | |
|---|---|---|---|---|---|---|---|---|---|---|
| | $Cr_{1/3}NbS_2$ | $[NH_4][Ni(HCOO)_3]$ | $[NH_4][Co(HCOO)_3]$ | | $[NH_4][Fe(HCOO)_3]$ | | $[NH_4][Mn(HCOO)_3]$ | | $KCo(HCOO)_3$ | |
| References | [25, 47] | [40] | [40] | | [40] | | [40] | | [41] | |
| **5NN** | | | | | | | | | | |
| d(M-M), (Å) | 10.653 | 11.830 | 11.888 | 11.766 – 11.910 | 11.935 | 11.796 – 11.978 | 12.020 | 11.857 – 12.063 | 11.498 | |
| $J5^1$ - $J5^6$ ($Å^{-1}$) | -0.0170 | -0.0063 | -0.0052 | -0.0037 to -0.0062 | -0.0043 | -0.0031 to -0.0064 | -0.0028 | -0.0011 to 0.0048 | -0.0039 | |
| ***c*NN** | | | | | | | | | | |
| d(M-M)=*c*(Å) | 12.097 | 8.021 | 8.190 | 8.224 | 8.318 | 8.365 | 8.489 | 8.537 | 8.469 | 6.888 |
| $J_c$ ($Å^{-1}$) | 0.0075 | 0.0049 | 0.0051 | 0.0049 | 0.0053 | 0.0051 | 0.0057 | 0.0055 | 0.0046 | -0.0004 |
| **6NN** | | | | | | | | | | |
| d(M-M) (Å) | 13.390 | 10.836 | 10.975 | 10.936 – 11.013 | 11.083 | 11.043 – 11.141 | 11.237 | 11.188 – 11.297 | 10.986 | |
| $J6^1$ - $J6^3$ ($Å^{-1}$) | -0.0413 | -0.0342 | -0.0341 | -0.0326 to -0.0361 | -0.0342 | -0.0320 to -0.0362 | -0.0340 | -0.0321 to -0.0363 | -0.0382 | |
| *j*(O1)-*j*(O6) [(d)] ($Å^{-1}$) | -0.0202 | -0.0153 | -0.0153 | -0.0149 to -0.0163 | -0.0153 | -0.0141 to -0.0166 | -0.0153 | -0.0140 to -0.0168 | -0.0173 | |
| MO1M[(e)]- MO6M[(e)] | 175.45° | 166.46º | 167.06º | 166.01º – 168.46º | 167.45 | 165.88º – 169.35 º | 168.07º | 159.59° – 170.23° | 169.41º | |
| $J6'^1$ - $J6'^3$($Å^{-1}$) | -0.0003 | 0.0012 | 0.0014 | 0.0008 – 0.0012 | 0.0018 | 0.0013 – 0.0016 | 0.0038 | 0.0017 – 0.0019 | -0.0006 | |

[a]By X-ray diffraction from single crystal (XDS).
[b] The refinement converged to the residual factor (*R*) values.
[c] *J*n<0 – AFM, *J*n>0 – FM
[d] *j* - maximal contributions of the intermediate X ion into the AFM component of the *J*n coupling
[(e)] $M_i$-X-$M_j$ bonding angl.

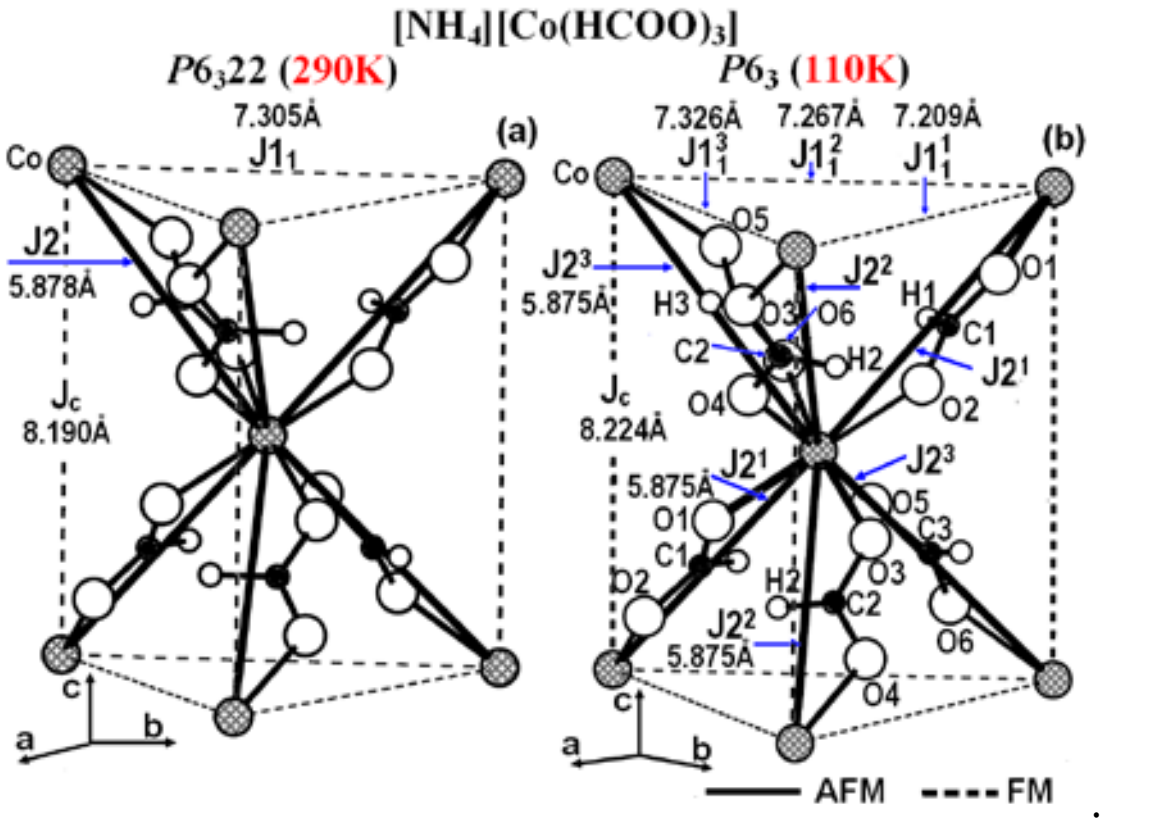

**Fig. 1** The local coordination environment of the $Co^{2+}$ ion with its six neighboring $Co^{2+}$ ions through bridging $HCOO^{-}$ in [NH4][Co(HCOO)3] crystallized in two space groups: $P6_322$ (290 K) (**a**) and $P6_3$ (110 K) (**b**).

## 2.3 Comparison of Crystal Structures of Metal Formates $[NH_4][M(HCOO)_3]$ ($M^{2+}$ = Mn, Fe, Co, Ni) and $KCo(HCOO)_3$ and Chiral Magnetic Soliton $Cr_{1/3}NbS_2$

The structural types of $NH_4Mn(HCOO)_3$ and $KCo(HCOO)_3$ are homeotypes, since they are very similar in mutual atomic locations in the crystal structure. From the first glance, these homeotypes crystal structure have nothing in common with that of $Cr_{1/3}NbS_2$, except the symmetry space group $P6_322$. The intercalation layered compound $Cr_{1/3}NbS_2$ is two-dimensional, while the crystal structure of divalent transition metal formats [40-42] is three-dimensional and chiral. In the crystal structure of $Cr_{1/3}NbS_2$, magnetic $Cr^{3+}$ atoms are located ordered in octahedral voids between S-Nb-S sandwich layers. $CrS_6$ octahedra are not linked to each other.

The crystal structure of metal formates consists of octahedral metal centers linked through *anti-anti* formate ligands (Fig. 1), while $NH_4^+$($K^+$) cations sit in the channels. The structural chirality in metal formates arises from the handedness imposed by the formate ligands around the metal ions and the presence of units with only one (*anti-anti*) handedness of three possible different O−C−O bridging modes: *syn*−*syn*, *syn*−*anti*, *anti*−*anti* [43-45].

In spite of apparent difference, magnetic ions have similar mutual location in these three structural types: $Cr_{1/3}NbS_2$, $NH_4Mn(HCOO)_3$, and $KCo(HCOO)_3$. In the space group $P6_322$, Cr and Mn ions occupy the same fixed positions (2c: 1/3, 2/3, 1/4) in the structural types $Cr_{1/3}NbS_2$ and $NH_4Mn(HCOO)_3$. In the structural type $KCo(HCOO)_3$, the position of Co ions (2d: 1/3, 2/3, 3/4) are different in just displacement by ½ along the *c* axis.

Besides, magnetic ions have the identical octahedral coordination ($MnO_6$, $CoO_6$, and $CrS_6$). The crystal lattices of magnetic $Cr^{3+}$ ions in $Cr_{1/3}NbS_2$ and $M^{2+}$ in the structural types $NH_4Mn(HCOO)_3$ and $KCo(HCOO)_3$ comprise flat triangular planes parallel to the *ab* plane (Fig. 2a, b). The triangular planes are located one above another at a distance of *c*/2 with displacements by *a*/3 and *b*/3. The crystal lattice of magnetic $M^{2+}$ ions, just like that of $Cr^{3+}$ ions, can be represented as two identical sublattices embedded into each other: the first one is formed by ions in trigonal prisms vertices, the other one – by ions in centers (Fig. 2c, e).

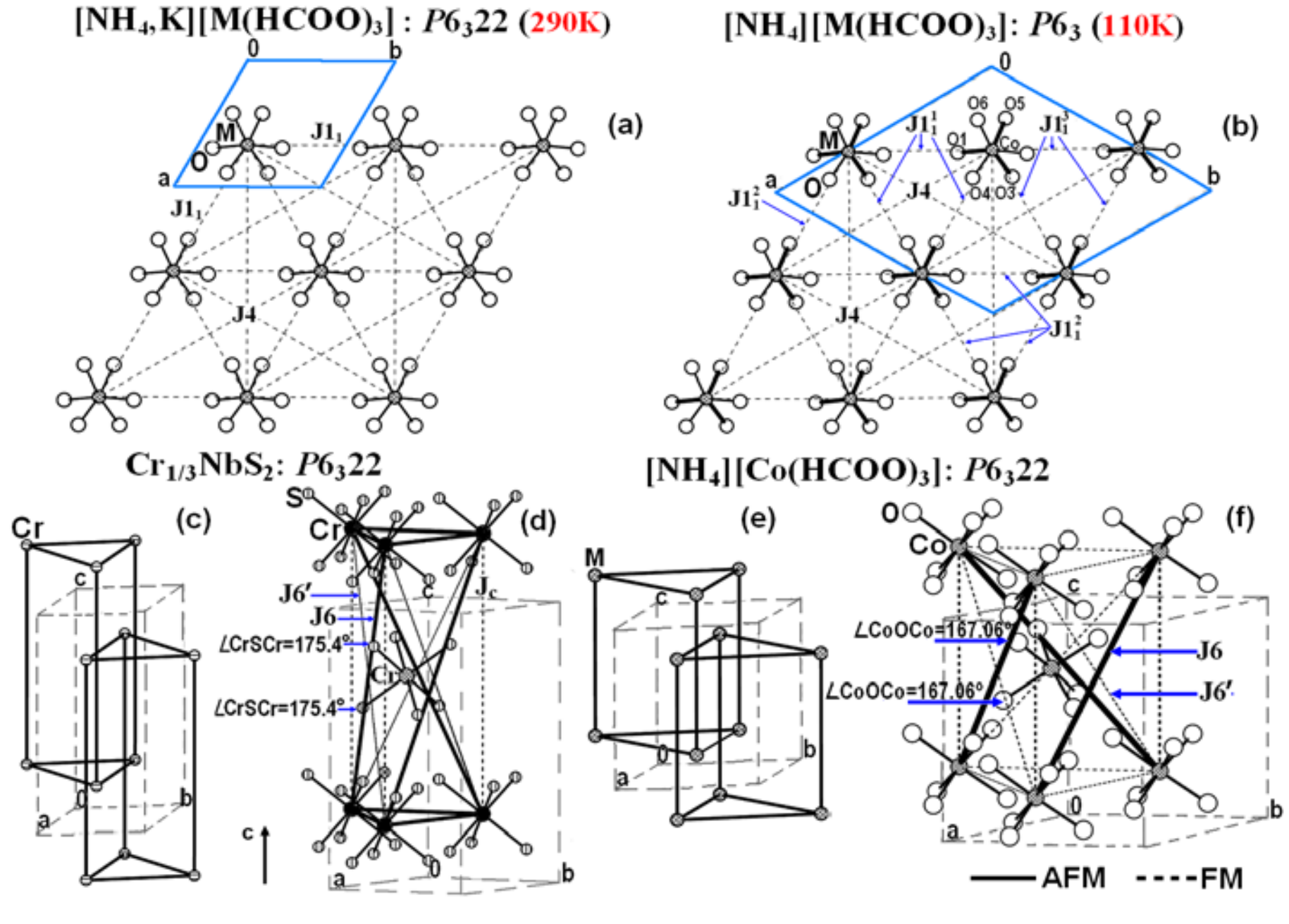

**Fig. 2** Triangular magnetic lattices with FM *J*1 and FM *J*4 couplings in metal formates $[NH_4][M(HCOO)_3]$ $P6_322$ (**a**) and $P6_3$ (**b**) phases. Elementary fragments – centered triangular prism $M_7$ and $J1_1$, *J*6, *J*6', and *J*c couplings in $Cr_{1/3}NbS_2$ (**c, d**) and $[NH_4][M(HCOO)_3]$ (**e, f**). On this and other figures, the line width indicates the strength of *Jn* couplings. AFM and FM couplings are shown by solid and dashed lines, respectively.

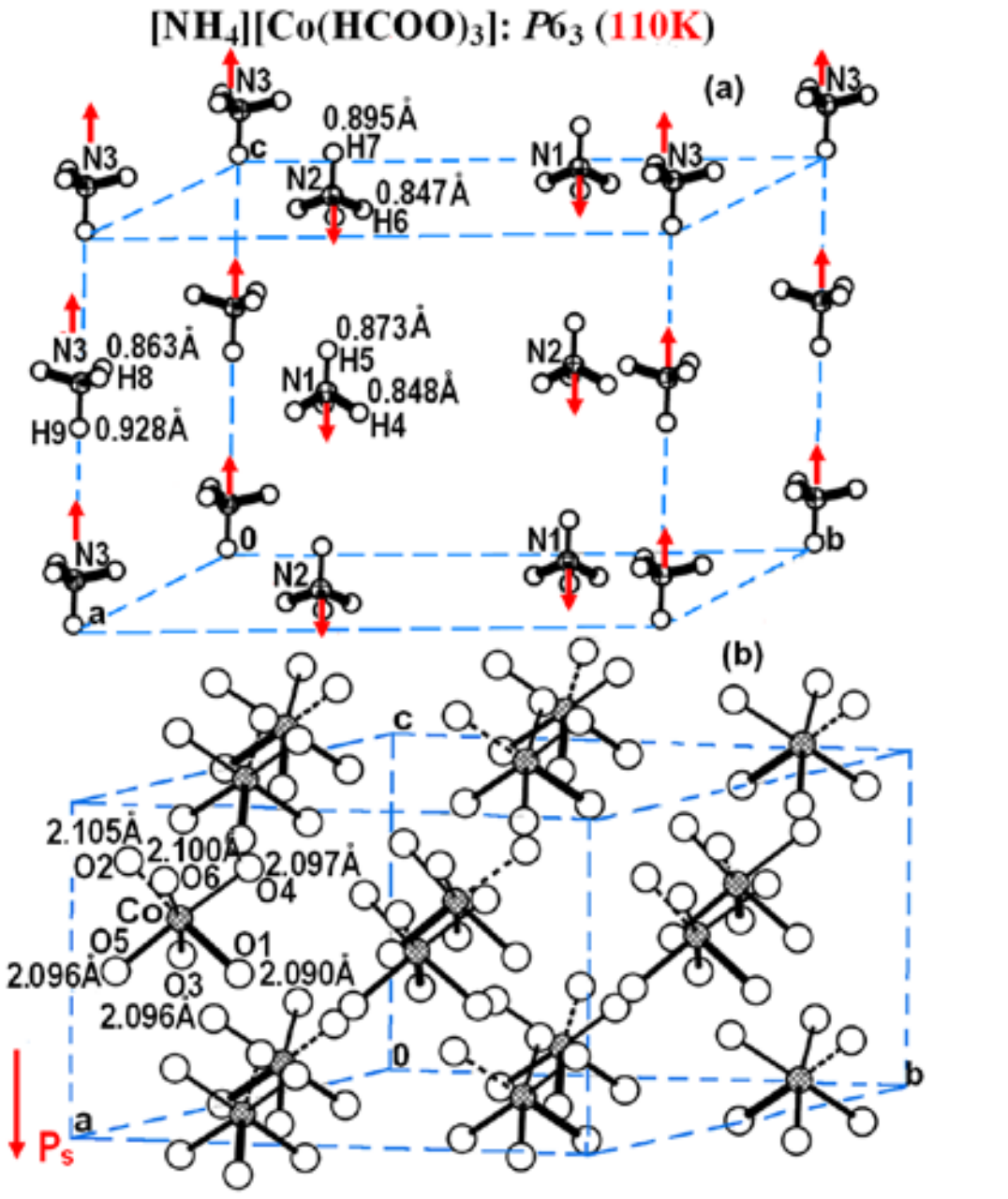

**Fig 3** The relation of N-H and Co-O bond lengths in $NH_4$-terahedra (**a**) and Co-O octahedra (**b**) in the *P*6$_3$ phase of $[NH_4][Co(HCOO)_3]$. Thick and thin (dashed) lines refer to short and long bonds, respectively. Arrows show the shifting direction of $N^{3-}$ in $NH_4$-tetrahedra.

The metal formates $[NH_4][M(HCOO)_3]$ ($M^{2+}$ = Mn, Fe, Co, Ni) undergo a transition from the *P*6$_3$22 (N182) to the *P*6$_3$ (N173) phase (Fig. 1b, 2b) [41]. Our analysis of the structural data [41] demonstrates that ammonium ions ordering (Fig. 3a) during such a transition is accompanied by just insignificant displacements in the magnetic sublattice, whose importance will be examined below, and slight distortion of $MO_6$ octahedra (Fig. 3b). As a result of such a transition, the length of M-O bonds to 3 oxygen atoms in the octahedron lower triangular face slightly decreases (down to 2.090-2.096 Å, 2.117-2.140 Å, and 2.168-2.186 Å for formate frameworks of Co, Fe, and Mn, respectively) as compared to the length of bonds to the opposite upper octahedron face (2.097-2.105 Å, 2.126-2.143 Å, and 2.183-2.194 Å for formate frameworks of Co, Fe and Mn, respectively). This effect of magnetic ion shifting along the 00-1 direction to one of the octahedron faces can be considered as weak electrical polarization. Note that the shortest and the longest M-O distances are in trans-positions in the octahedron, and this diagonal direction turns along the *c* axis from octahedron to octahedron (Fig. 3b). Since ordering of ammonium cations does not produce substantial structural effects in the metal-formate framework $[M(HCOO)_3]$, one can state that the structural type of new phases marked as 'the structural type $NH_4Mn(HCOO)_3$(*P*6$_3$)' is a homeotype of the structural type $[NH_4][M(HCOO)_3]$ (Figs. 4 and 5).

It turns out that the crystal structure of magnetic sublattices and the location of intermediate ions ($S^{2-}$($O^{2-}$))

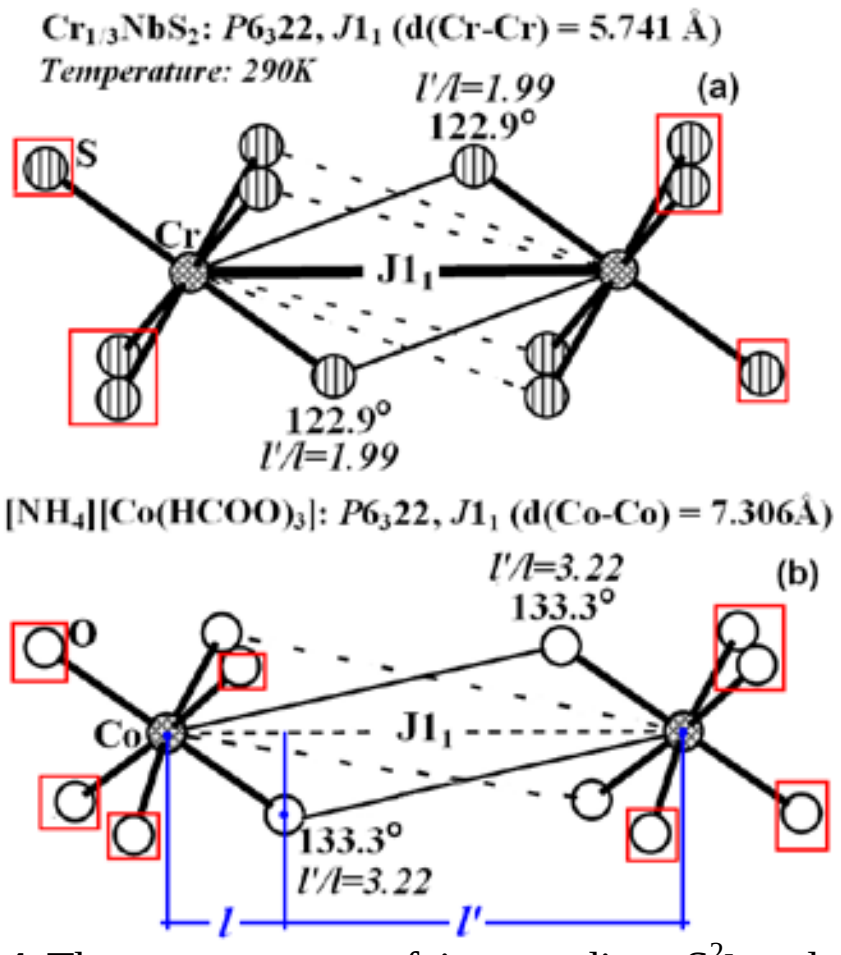

**Fig. 4** The arrangement of intermediate $S^{2-}$ and $O^{2-}$ ions in local space of $J1_1$ couplings in $Cr_{1/3}NbS_2$ (**a**) and $[NH_4][M(HCOO)_3]$ (**b**).

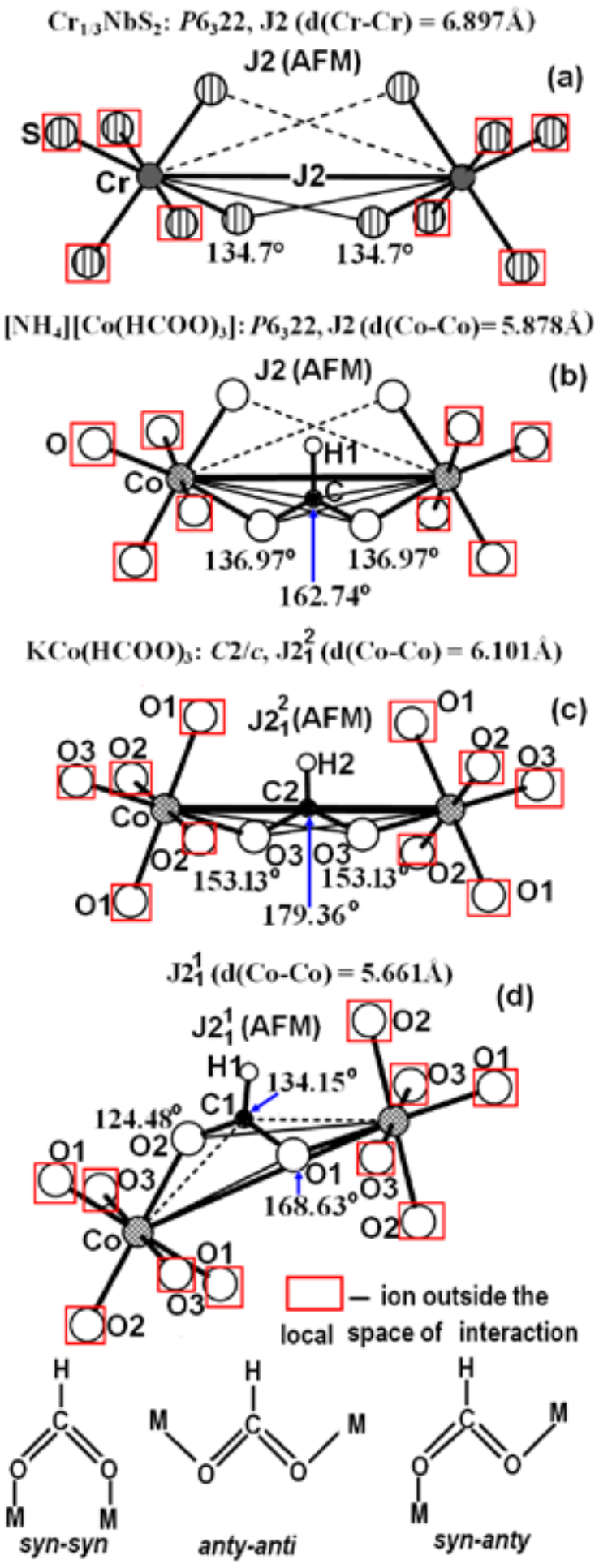

**Fig. 5** The arrangement of intermediate $S^{2-}$ and $O^{2-}$ ions in local space of AFM *J*2 (*anti-anti*) couplings in the *P*6$_3$22 phase of $Cr_{1/3}NbS_2$ (**a**), $[NH_4][Co(HCOO)_3]$ (**b**), and in AFM $J2_1^2$ (*anti-anti*) (**c**) and $J2_1^1$ (*syn-anti*) (**d**) in the *C*2/*c* phase of phase $KCo(HCOO)_3$.

between magnetic ions ($Cr^{3+}$($M^{2+}$)) determining the parameters of exchange magnetic couplings differ insignificantly in the structural types $NH_4Mn(HCOO)_3$, $NH_4Mn(HCOO)_3$($P6_3$), and $KCo(HCOO)_3$ and are very similar to respective characteristics in the structure of $Cr_{1/3}NbS_2$. However, there are two important differences of metal formates from $Cr_{1/3}NbS_2$ affecting the formation of magnetic couplings characteristics. The first one is concerned with $C^{2+}$ ions of bridging $HCOO^-$ groups additionally entering the local space of *J*2 (Fig. 1) and *J*5 (Fig. 6a, b) magnetic couplings between neighboring layers in the sublattice of magnetic $M^{2+}$ ions of metal formates. The second one consists in substantial difference of sizes of magnetic and intermediate ions in these compounds (the radii ratio $r_{M2+}/r_{Cr3+}$ = 1.12, 1.21, 1.27, and 1.35 Å for $Ni^{2+}$, $Co^{2+}$, $Fe^{2+}$, and $Mn^{2+}$, respectively, and $r_{O2-}/r_{S2-}$= 0.76). The following section will describe the result of integrated effect of the above factors on formation of the magnetic structure of the metal formats under study.

### 2.4 Magnetic Properties Metal Formates $[NH_4][M(HCOO)_3]$ ($M^{2+}$ = Mn, Fe, Co, Ni) and $KCo(HCOO)_3$

Although magnetic properties of the discussed metal-formate frameworks have been studied rather comprehensively, however nobody has yet been studied the shape of magnetic lattices of these compounds at changes of the strength of the external magnetic field, as was done in experiments with the chiral magnet $Cr_{1/3}NbS_2$ [8].

On the other hand, G.-Ch. Xu et al. [41] showed that magnetic and electric orderings coexisted in the magnetic members of the family of $[NH_4][M(HCOO)_3]$, and this could be expected for a new class of metal-organic multiferroics.

The above conclusion was grounded by the fact that these materials underwent a ferroelectric phase transition from the $P6_322$ to the $P6_3$ phase between 191 and 254 K. This transition is associated with ordering of ammonium cations and their displacement within the framework channels combined with spin-canted AFM ordering within 8-30 K. The formation of 6 polar $NH_4$ tetrahedra occurs in three channels (two per channel) within the unit cell, four in the same direction (00-1) and two in the opposite one (001) (Fig. 3a). Since the dipole moments of these tetrahedra do not compensate each other, there exists a polarization with the *c* axis as the polarization one.

According to refs. [40–42], metal formate frameworks display weak ferromagnetism in the low-temperature region. The ferroelectric nature of the structural phase transitions in ammonium metal formates was corroborated [46] by high-resolution micro-Brillouin scattering. The main role of ammonium cations in the mechanism of phase transition from the $P6_322$ to the $P6_3$ space group is also indicated by IR and Raman studies of $[NH_4][M(HCOO)_3]$ [47] and the absence of this phase transition in the hexagonal formate $K[Co(HCOO)_3]$, a homeotype of $[NH_4][M(HCOO)_3]$. However, the emerged coexistence of magnetic and electric orderings in metal-formate based on disorder-order transitions of $NH_4^+$ cations requires extra proofs, since the spin-canted AFM ordering ($T_N$ = 8.3 K) was also found [42] in the hexagonal polymorph $K[Co(HCOO)_3]$, despite the absence of $NH_4^+$ ions.

For $KCo(HCOO)_3$, the authors of Ref. 42 described the rare observation of an irreversible transformation from a chiral (space-group $P6_322$) to an achiral (monoclinic space group *C* 2/*c*) crystal through the syntheses of two polymorphs of $KCo(HCOO)_3$. It was established that the $P6_322$ phase of $KCo(HCOO)_3$ was the kinetic metastable phase, while the *C*2/*c* phase was the thermodynamic stable one. In the solid state, the hexagonal polymorph is irreversibly transformed to the monoclinic polymorph very slowly at ambient conditions. The solid state phase transition could be completed in 1 month at room temperature. However, when the sample of the $P6_322$ phase was kept at -18 °C in a refrigerator, the transformation became very slow. It was accompanied by an increase of the crystal density and a change of the coordination mode of some formates from *anti-anti* to *syn-anti* (Fig. 5c, d). The achiral (monoclinic) polymorph is an antiferromagnet ($T_N$ = 2.0 K).

## 3 Results and Discussion

### 3.1 Chiral Polarization of the Magnetic System of Hexagonal Metal Formates $[NH_4][M(HCOO)_3]$ with $M^{2+}$ = Mn, Fe, Co, Ni, and $KCo(HCOO)_3$

With respect to similarities and differences of magnetic structures of $[NH_4][M(HCOO)_3]$, $KCo(HCOO)_3$, and $Cr_{1/3}NbS_2$ caused by their crystal structures and compositions, we will examine the crystal lattice of magnetic M ions in these compounds as two identical sublattices embedded into each other: the first one is formed by ions in trigonal prisms vertices and the other one by ions in centers (Fig. 2c, e). In the lattice, one can single out elementary fragments in the form of triangular centered $M_7$ prism (Fig. 2d, f).

According to our calculations (Table 2), the crystal structures of both $P6_322$ and $P6_3$ phases of metal formate frameworks cause as early as at room temperature the existence of weak $J1_1$ ($J1_1$ = -0.0002–0.0018 Å$^{-1}$, d(M-M) = 7.209–7.399 Å) couplings along triangles sides in triangular planes parallel to the *ab* plane (Fig. 2a, b and 4; Table 2). This ensures the quasi-one-dimensional character of the metal formate magnetic system, whereas the emergence of quasi-one-dimensional structure in $Cr_{1/3}NbS_2$ requires extra displacements of $S^{2-}$ ions that are possible upon temperature reduction. According to our calculations [17], at room temperature both couplings (AFM $J6$ and AFM $J1_1$) in $Cr_{1/3}NbS_2$ are strong ($J6/J1_1$ = 0.77). However, the AFM $J1_1$ coupling is unstable. Two $S^{2-}$ ions (Cr-S-Cr angle 122.9°) making a large contribution to the AFM component of this coupling are located near the boundary ($l'/l$ = 1.99) of the central one-third of the local space between magnetic $Cr^{3+}$ ions (Fig. 4a) and, therefore, near the critical position “c” ($l'_n/l_n$ = 2) [27, 28]. An insignificant displacement (by 0.006 Å along the *a* or *b* axis from the center in parallel to the Cr-Cr bond line) of $S^{2-}$ ions at the temperature decrease induces a dramatic decrease of the strength of AFM $J1_1$ couplings ($J6/J1_1$ = 10.07), weakens the frustration, and transforms the magnetic system of $Cr_{1/3}NbS_2$ into a quasi-one-dimensional one. The triangular planes of both $P6_322$ and $P6_3$ phases of metal-formate frameworks contain two more ($J1_2$ and $J4$) couplings (Fig. 2a, b)), which are ferromagnetic and do not compete with FM $J1_1$ couplings. The next-nearest-neighbor $J1_2$ couplings in linear chains along the triangles sides are three- to fivefold weaker than the $J6$ couplings, while the $J4$ couplings are as weak as the $J1_1$ ones. Thus, all the couplings in triangular planes of metal formates are weak ferromagnetic.

Ferromagnetic orientation of spins in each triangular plane and antiferromagnetic one of neighboring planes (in opposite directions) in both $P6_322$ and $P6_3$ phases are predetermined by antiferromagnetic $J2$, $J3$, and $J5$ couplings (Table 2) between neighboring planes (Fig. 5 and 6a, b).

The strength of AFM $J2$ couplings between the nearest neighboring $M^{2+}$ ions is significantly higher than that of respective couplings in $Cr_{1/3}NbS_2$. The latter occurs due to extra entry of the carbon atom from the bridging $HCOO^-$ group, which makes a large contribution to the interaction AFM components into the central one-third of the local space of the $J2$ coupling (Fig. 5a-c). This contribution is especially large for metal formates of $Ni^{2+}$ and $Co^{2+}$ ($J6/J2$ = 1.4–1.6) with small magnetic ion radii, but it decreases along with the magnetic ion size increase ($J6/J2$ = 1.9–3.3 for $Fe^{2+}$ and $Mn^{2+}$) (Table 2). Note that, aside from the location of the intermediate ion in the local space, its contribution into the formation of magnetic coupling is determined by this ion size. For the calculation, we selected the radius of the carbon ion equal to its covalent radius in $sp^2$ hybridization. Unfortunately, we have not managed to find in the literature experimental data that could be compared to those obtained using our method, in order to estimate the correctness of this radius use. If one decreases the size of the carbon atom, its contribution to the AFM component of the $J2$ coupling will decrease accordingly. The interplane $J3$ and $J5$ couplings in metal formates ($J6/J3$ = 9.09–15.5 and $J6/J5$ = 5.4–29.2) are substantially weaker than in the soliton $Cr_{1/3}NbS_2$ ($J6/J3$ = -2.5 and $J6/J5$ = 2.4) because of the difference in sizes of intermediate ions of oxygen and sulfur. Moreover, a larger size of $S^{2-}$ ions in comparison to $O^{2-}$ ions allows extra entry of the former to the local sphere of the $J3$ coupling in $Cr_{1/3}NbS_2$, make a substantial contribution to its FM component, and transform this coupling into a ferromagnetic one, while in metal-formates it is antiferromagnetic. One should emphasize that crystallographic equivalents of interplane $J2$, $J3$, $J5$, and $J_c$ couplings in the lattice of magnetic ions are at the same time magnetic equivalents and do not form chiral helices along the *c* axis. Thus, the magnetic structure of metal formate frameworks in both phases ($P6_322$ and $P6_3$) comprises a system of two sublattices with opposite spins (Fig. 6), if one takes into account FM $J1$ and $J4$ couplings in triangular planes and AFM $J2$, $J3$, and $J5$ couplings between neighboring planes.

Competition is here provided by dominating AFM left-handed $J6$ helices along the *c* axis (Table 2). The left-handed spin helices in both $P6_322$ and $P6_3$ phases of metal formate frameworks, just like in $Cr_{1/3}NbS_2$, are formed from dominating in force couplings between triangular planes of magnetic ions along just one of two crystallographically equivalent diagonals of side faces embedded into each other trigonal prisms $M_7$ (Fig. 2c-f) composing crystal lattices of magnetic ions in the structural types $NH_4Mn(HCOO)_3$, $NH_4Mn(HCOO)_3(P6_3)$, and $KCo(HCOO)_3$. They comprise intertwined left- handed AFM helices twisted along the *c* axis (Fig. 7a-e).

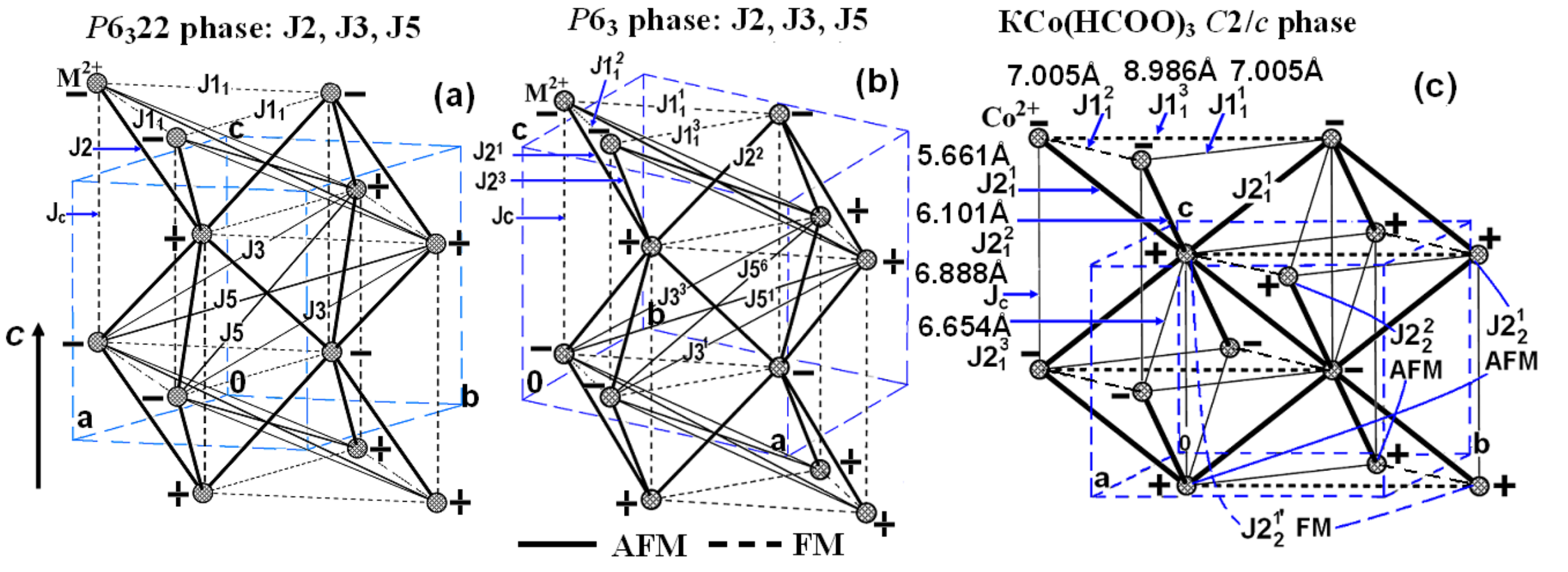

**Fig. 6** AFM $J2$, $J3$, and $J5$ couplings between neighboring lattices in $P6_322$ (**a**), $P6_3$ (**b**) and $C2/c$ (**c**) phases of metal-formate frameworks.

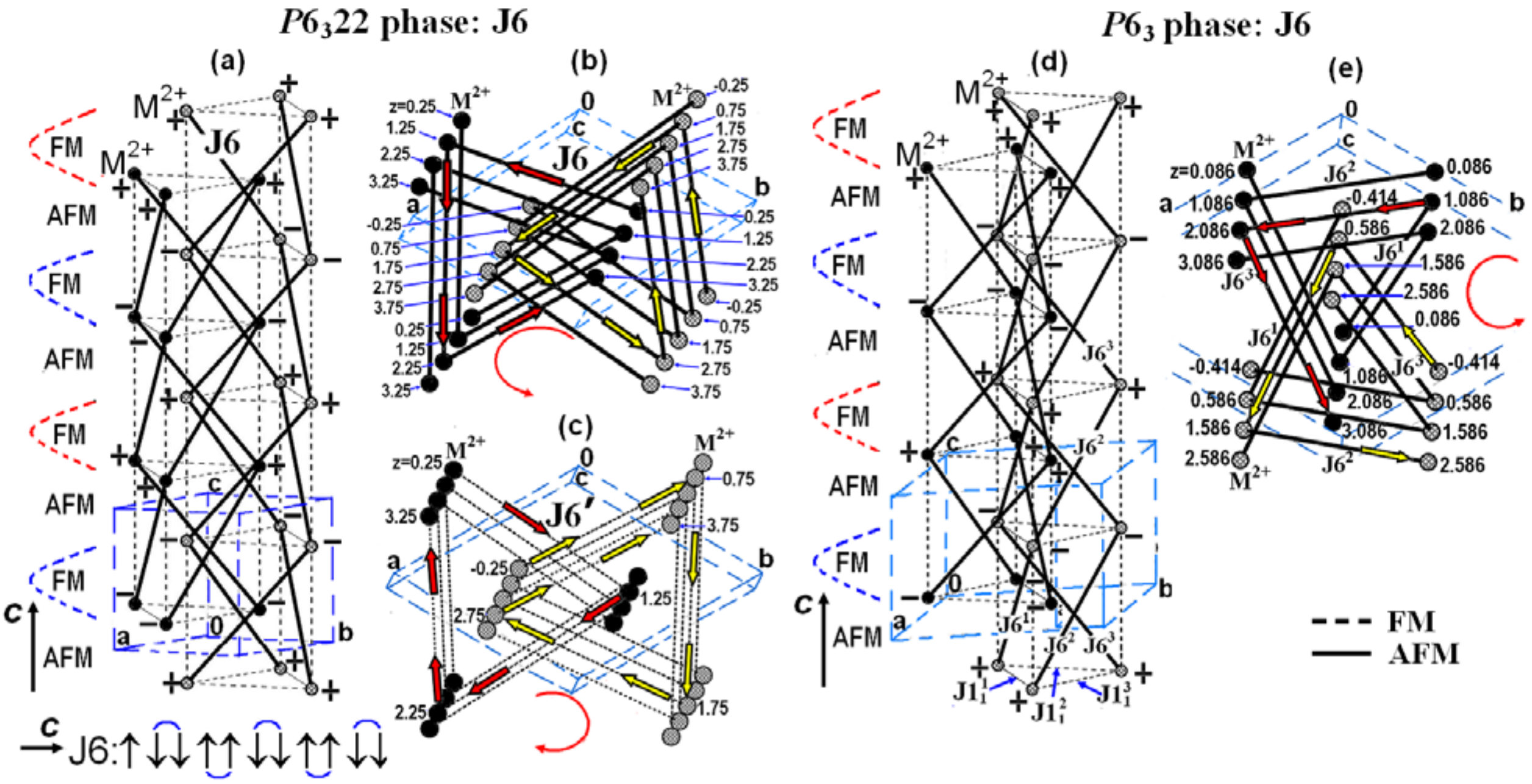

**Fig. 7** Left-handed *J*6 helices: along the *c* axis (**a**) and the view along [001] (**b**) and right-handed *J*6' helices: the view along [001] (**c**) in the $P6_322$ фазе. Left-handed *J*6 helices: along the *c* axis (**d**) and the view along [001] (**e**) in the $P6_3$ phase.

**Table 3** Sum of forces of interplane couplings (∑J*) orienting magnetic moments of individual planes in ($[NH_4][M(HCOO)_3]$, $KCo(HCOO)_3$, and $Cr_{1/3}NbS_2$

| $M^{2+}$ | Ni | Co | | Fe | | Mn | | KCo | $Cr_{1/3}NbS_2$ |
|---|---|---|---|---|---|---|---|---|---|
| Space Group | $P6_322$ | $P6_322$ | $P6_3$ | $P6_322$ | $P6_3$ | $P6_322$ | $P6_3$ | $P6_322$ | $P6_322$ |
| 1 ∑*J**, $Å^{-1}$ | -0.075 | -0.070 | -0.069 | -0.065 | -0.064 | -0.058 | -0.057 | -0.072 | -0.069 |
| 2 ∑*J**, $Å^{-1}$ | 0.008 | 0.002 | 0.0004 | -0.003 | -0.004 | -0.010 | -0.011 | -0.005 | 0.020 |
| 3 ∑*J**, $Å^{-1}$ | -0.007 | -0.002 | -0.0004 | 0.003 | 0.004 | 0.010 | 0.011 | 0.005 | 0.014 |
| 4 ∑*J**, $Å^{-1}$ | 0.075 | 0.070 | 0.069 | 0.065 | 0.064 | 0.058 | 0.057 | 0.072 | 0.102 |
| 5 ∑*J**, $Å^{-1}$ | -0.075 | -0.070 | -0.069 | -0.065 | -0.064 | -0.058 | -0.057 | -0.072 | -0.069 |

Contributions to the AFM component of the *J*6 coupling emerge under the effect of two $O^{2-}$ ions (M-O-M angles 166–170°) forming the edge of the octahedron of the magnetic ion centering the trigonal prism $M_7$ (Fig. 2f). Note that carbon atoms are not included into the local sphere of *J*6 couplings and, therefore, do not participate in their formation. Figure 2d, f show the similarity of *J*6 couplings formation in metal formates and $Cr_{1/3}NbS_2$.

It is important to emphasize that the direction of magnetic moments (spins) of adjacent atoms in this left-handed spiral chain cannot be perfectly antiferromagnetic. They are slightly canted (Fig. 8).
The reason of chiral polarization of the magnetic system is a sharp nonequivalence (|*J*6/*J*6'| = 9–64) in the strength of crystallographically equivalent left-handed *J*6 (Fig. 7b) and right-handed *J*6' (Fig. 7c) magnetic helices due to the noncentrosymmetric location of $O^{2-}$ ($S^{2-}$ в $Cr_{1/3}NbS_2$) ions (Table 2). The magnetic *J*6' coupling along another crystallographically equivalent diagonal of the prism side face is significantly weaker than the AFM *J*6 coupling, since $O^{2-}$ ions do not enter the central one-third of its local space (Fig. 2f). Moreover, the *J*6' is a weak ferromagnetic one in all metal formates, except

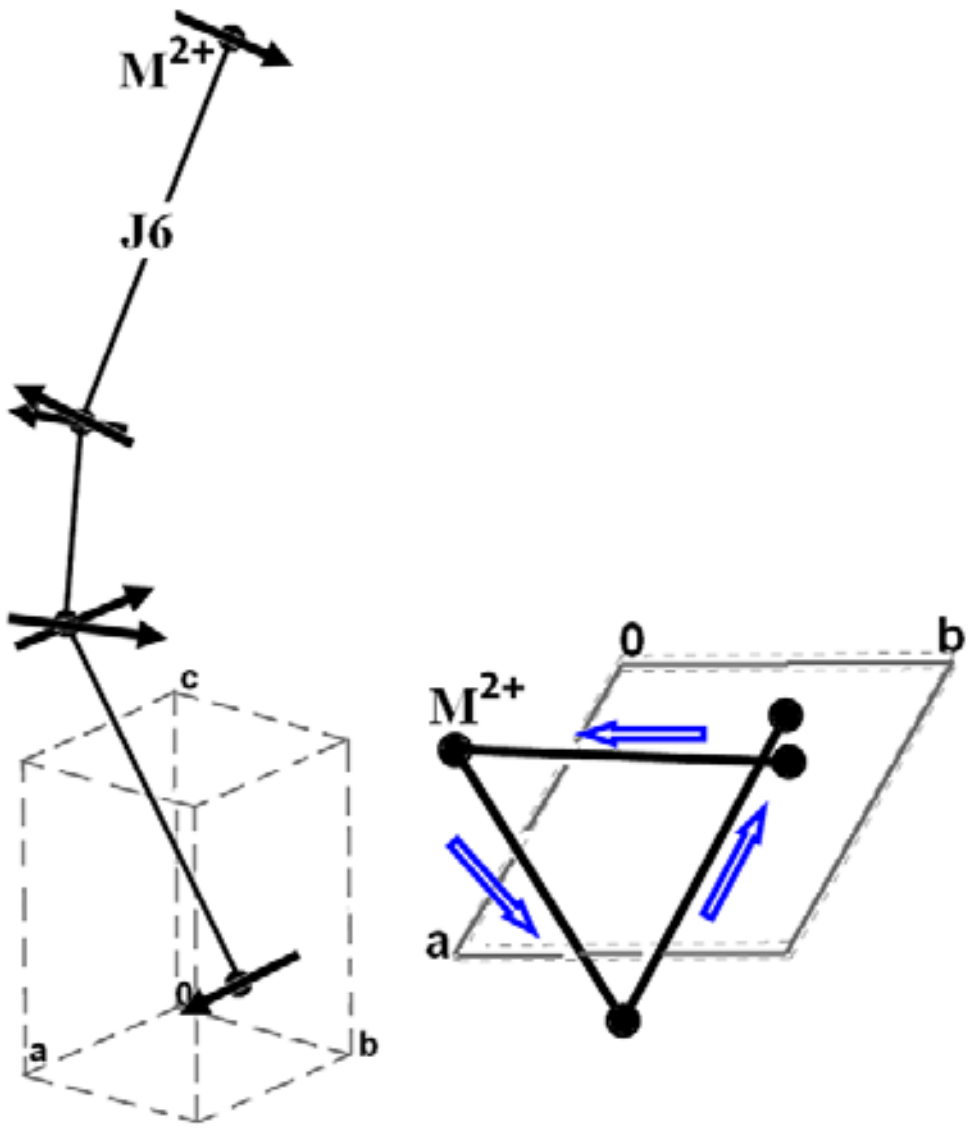

**Fig. 8.** Non-collinearity of magnetic moments in the left-handed AFM spin spiral (a) and the general view along the [001] left-handed spiral of magnetic ions.

$KCo(HCOO)_3$, in which it is weak antiferromagnetic

To sum up, we obtain the following picture of ordering of FM triangular planes along the *c* axis under the effect of strong helical AFM *J*6 couplings. They turn spins in sublattices to the opposite direction and double the magnetic lattice *c* parameter. As a result, the magnetic ordering of FM triangular planes along the *c* axis becomes like↑↑↓↓; i.e., each block of two neighboring triangular planes is oriented ferromagnetically, while blocks themselves are ordered according to the AFM type (Fig. 7a, d).

If one formally combines the strengths of interplane *J*2, *J*5, *J*3, $J_c$, and *J*6 couplings taking into account their sign (spin direction + or -) in respective plane and multiplicities (multiplicities of *J*2, *J*3, and *J*6 are equal to 6, those of *J*5 and $J_c$ are equal to 12 and 2, respectively), their sum $\sum J^* = J^*2 + 2J^*5 + J^*3 + J^*_c/3 + J^*6$) will change along the *c* axis between planes. During the $\sum J^*$ calculation, the *J*6' value was not taken into account due to its small value, whereas the average value of a certain type of $J_n$ couplings was used for the $P6_3$ phase. Changes in the sum of interplane couplings ($\sum J^*$) orienting magnetic moments of individual planes along the *c* axis are shown in Table 3. The obtained results indicate that in metal-formates of $[NH_4][M(HCOO)_3]$ with $M^{2+}$ = Mn and Fe and $KCo(HCOO)_3$ one observes the strongest effect of chiral *J*6 couplings, since after combining all the couplings ordering of triangular FM planes along the *c* axis is preserved in them in the form ↑↑↓↓. Based on the above, one could assume that the probability of emergence of solitons in them was higher than that in metal formate with $M^{2+}$ = $Ni^{2+}$ and $Co^{2+}$, in which after combining, ordering of planes along the *c* axis acquires the form:↑↓↑↓. This occurs because of the increased strength of *J*2 couplings in these compounds, and the combined effect of *J*2, *J*3, and *J*5 couplings between neighboring planes effectively competes with those in chiral helices. However, as was shown above, if one uses a smaller radius of the carbon atom in *J*2 couplings calculation, its contribution to the AFM component of the *J*2 coupling decreases accordingly, and, therefore, its competition with the chiral *J*6 coupling becomes weaker. Thus, changes in the sum of strengths of interplane couplings ($\sum J^*$) orienting magnetic moments along the *c* axis shows that the direction of spins, while remaining the same for ions of the same plane, changes between planes with a periodicity in four planes in all the examined metal formates and $Cr_{1/3}NbS_2$. In spite of the formal character of this approach, it illustrates in a simplified form the possibility of origination of a superstructure, which, according to Dzyaloshinskii [48], could emerge in a system composed of two sublattices with opposite spins and superposition of pulsation (in this case in the form of *J*6 couplings) having a large period. Under the effect of relativistic forces, the value of the superstructure period can be multiply increased. We suggest that the Dzyaloshinskii–Moriya interaction could result in ordering in the form of a chiral magnetic soliton lattice of non-collinear left-handed AFM spin helices competing with AFM interhelix spin couplings.

### 3.2 Interrelation Between Chiral Polarization of Crystalline and Magnetic Structures in $KCo(HCOO)_3$

In the course of transition of $KCo(HCOO)_3$ [42] from the metastable noncentrosymmetric hexagonal $P6_322$ phase to the stable centrosymmetric monoclinic *C*2/*c* phase, one observes the displacement of triangular planes of $Co^{2+}$ ions located in parallel to the *ab* plane relatively to each other (Figs. 2a, 9a–d and 10a). In this case, preservation of the local coordination environment of the $Co^{2+}$ ion with its six neighboring $Co^{2+}$ ions through bridging $HCOO^-$ ligands is possible only at the decrease of the distance between planes, which results in the decrease of the parameter *c* by 1.54 Å and strong distortion of the local environment. In the chiral $P6_322$ phase in the local coordination environment of $Co^{2+}$ ions comprising a trigonal prism (d(Co-Co) = 5.843 Å), all bridging $HCOO^-$ groups are in *anti-anti* mode (Fig. 9a,b). The transformation to an achiral *C*2/*c* phase (Fig. 9c,d) is accompanied by a change of the coordination mode of four bridging $HCOO^-$groups from *anti-anti* to *syn-anti* (d(Co-Co) = 5.661 Å, Fig. 5d), while two $HCOO^-$ groups remain in *anti-anti* (d(Co-Co) = 6.101 Å, Fig. 5c) mode.

This transformation from chiral to achiral polymorph of $K[Co(HCOO)_3]$ does not virtually affect the parameters of magnetic couplings in planes parallel to the *ab* plane (Figs. 2a and 10a; Table 1). All the couplings ($J1_1^2$, $J1_1^3(J_b)$, $J4^1$, $J4^2$, and $J4^2$) remain weak FM ones, except the $J1_1^1$ coupling along one of the triangle sides, which, still remaining very weak, is transformed into the AFM state and contributes some competition into the triangular plane. Interplane couplings undergo substantial changes.

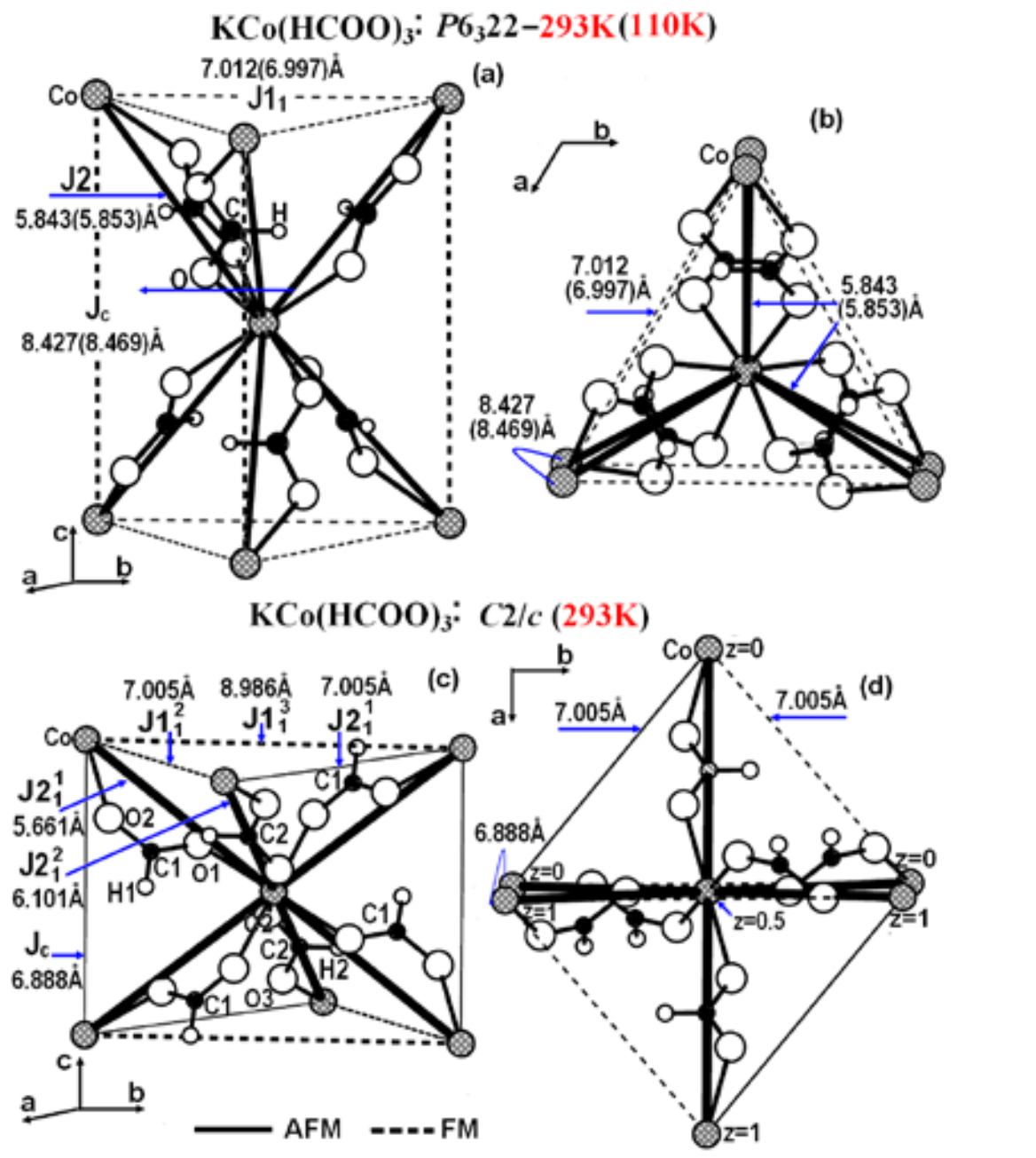

**Fig. 9** The local coordination environment of the $Co^{2+}$ ion with its six neighboring $Co^{2+}$ ions through bridging $HCOO^-$ in two polymorphs of $KCo(HCOO)_3$: chiral (**a**) and achiral (**c**). A view along [001] (**b**) and (**d**).

As was shown above, in the quasi- one-dimensional chiral polymorph of $KCo(HCOO)_3$, the AFM *J6* couplings (*J6* = -0.0382 Å $^{-1}$, d(M-M) = 10.98Å) between triangular planes of magnetic ions through the planes formed by left-handed spin helices along the *c* axis (Fig. 7a, b) were dominating in strength. In the achiral polymorph, the $J2_1^1$ couplings ($J2_1^1$ = -0.0538 Å$^{-1}$, d(Co-Co) = 5.661 Å) between neighboring planes through *syn-anti* bridging $HCOO^-$ groups (Fig. 5d) and $J2_1^2$ ones ($J2_1^2$ = -0.0581 Å $^{-1}$, d(Co-Co) = 6.101Å) through *anti-anti* bridging $HCOO^-$ groups (Fig. 5c) are dominating, which transforms the polymorph magnetic structure into the achiral three-dimensional AFM one (Figs. 6c and 10b-c). A substantial AFM contribution to $J2_1^1$ couplings (Fig. 5d) is provided by the O1 ion ($j_{O1}$ = -0.0814 Å $^{-1}$, Co-O1-Co angle 168.63º) from the *syn-anti* bridging $HCOO^-$ group, whereas the C1 ion, in opposite, contributes to the FM component of this interaction ($j_{C1}$ = 0.0291 Å $^{-1}$, Co-C1-Co angle134.15º), thus reducing the strength of the AFM coupling. The antiferromagnetic $J2_1^2$ coupling (Fig. 5c) is formed under the effect of large AFM contributions of C2 and two O3 ions from *anti-anti* bridging $HCOO^-$ group ($j_{C2}$ = -0.0383 Å$^{-1}$, Co-C2-Co angle 179.36º, $j_{O3}$ = -0.0099 Å $^{-1}$, Co-O3-Co angle 153.13º). However, these nearest-neighbor AFM $J2_1^1$ and $J2_1^2$ couplings are frustrated, since they compete with the next-nearest-neighbor along linear chains by $J2_2^1$ ($J2_2^1/J2_1^1$ = 0.33 – 0.37, d(Co-Co) = 11.322Å) and $J2_2^2$ ($J2_2^2/J2_1^2$ = 0.70, d(Co-Co) = 12.202Å) couplings ( Fig. 10b, c) that are comparatively strong AFM ones.

To sum up, we demonstrated similarity and differences of metastable and stable phases of $K[Co(HCOO)_3]$ marking the features of magnetic and crystalline phases that could indicate to the possibility of the reversible transition. Duan et al. [42] described the crystal structure of the *C*2/*c* phase from another point – without emphasizing its common features with the metastable crystalline structure. They based the analysis on the planes parallel to the *bc* plane, in which $Co^{2+}$ ions at shortest distances d(Co-Co) = 5.661Å are bound by *syn-anti* bridging $HCOO^-$ group into a tetragon. The above planes are bound to each other at short distances d(Co-Co) = 6.101Å by the *anti-anti* bridging $HCOO^-$ groups. Figure 9b,c show AFM $J2_1^1$, $J2_2^1$, and *J3* couplings in the *bc* plane and AFM $J2_1^2$, and $J2_2^2$ couplings between planes along the *a* axis.

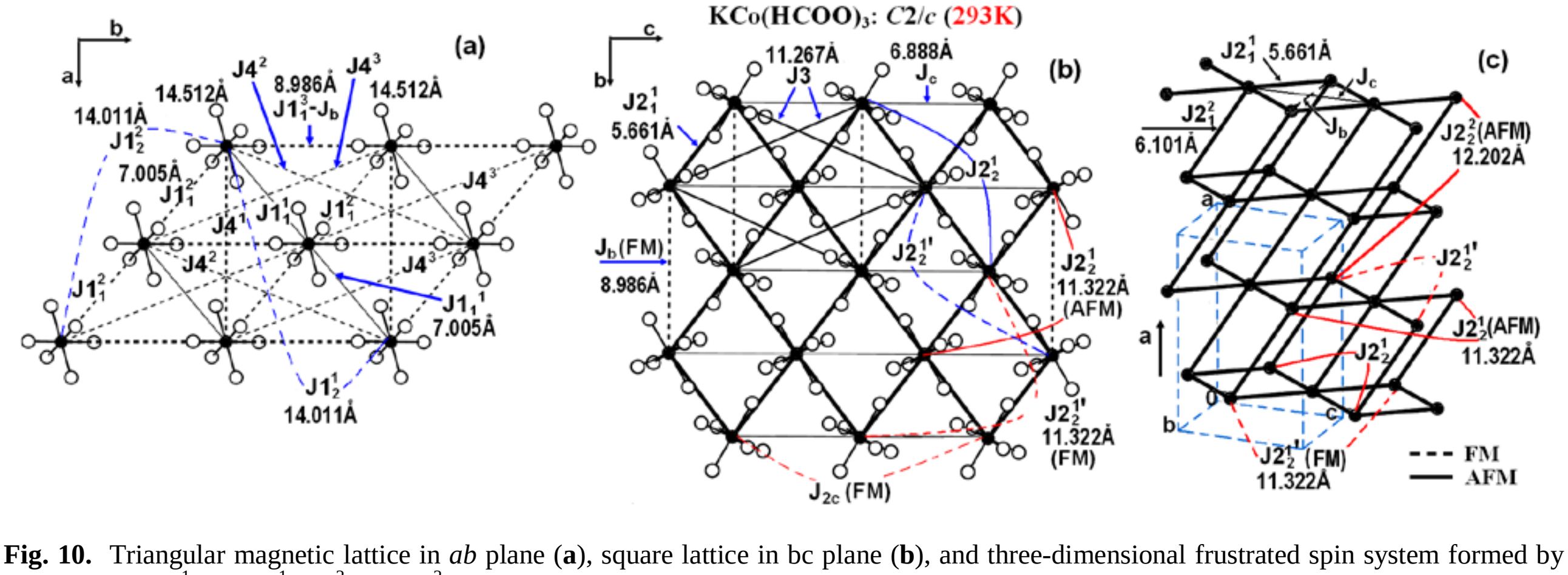

**Fig. 10.** Triangular magnetic lattice in *ab* plane (**a**), square lattice in bc plane (**b**), and three-dimensional frustrated spin system formed by strong AFM $J2_1^1$ and $J2_2^1$, $J2_1^2$ and $J2_2^2$ couplings (**c**) in the *C*2/*c* phase of $KCo(HCOO)_3$.

If the metastable $P6_322$ $KCo(HCOO)_3$ phase having a chiral crystal structure and, according to our calculations, comprising a chiral magnetic soliton is used as a basis, then the family of metal-formates $[NH_4][M(HCOO)_3]$ ($M^{2+}$ = Mn, Fe, Co, Ni) and $KCo(HCOO)_3$ becomes of special interest to a researcher. In view of this, the application of the external field or substitution of the $K^+$ ion by the $NH_4^+$ ion can be considered as stabilization [49, 50] of the chiral magnetic soliton and the chirality of the crystal structure (*anti-anti* coordination mode of formates).

## 4 Conclusions and Summary

The search of potential chiral magnetic solitons similar to $Cr_{1/3}NbS_2$ in the ICSD database has been performed among magnetic compounds crystallizing in the non-centrosymmetric hexagonal space group $P6_322$ using the earlier developed crystal chemistry method for determination of magnetic interaction parameters on the basis of structural data. The crystal chemistry criteria of the search included the following characteristics: (1) absence of the symmetry center, (2) presence of chiral spin helices – bases of the lattice, and (3) domination the chiral spin helices in the system and their competition with other interactions promoting the formation of superstructures with a large period. Five compounds corresponding to these criteria have been identified: metal formates $[NH_4][M(HCOO)_3]$ with $M^{2+}$ = Mn, Fe, Co, Ni and $KCo(HCOO)_3$.

According to our calculations, the magnetic structure of these metal formates, just like that of the chiral magnetic soliton $Cr_{1/3}NbS_2$, is formed by dominating left-handed spin helices between triangular planes of magnetic ions through the plane of just one of two crystallographically equivalent diagonals of side faces of embedded into each other trigonal prisms building up the crystal lattice of magnetic ions. These helices are oriented along the *c* axis and packed into two-dimensional triangular lattices in planes perpendicular to these helices directions and lay one upon each other with a displacement. The magnetic ordering of triangular planes along the *c* axis under effect of strong helical AFM *J*6 couplings could have the form: ↑↑↓↓. However, non-collinearity of the AFM ordering of AFM left-handed spin helices and their competition with AFM interactions between adjacent planes in the structure indicate to the possibility of nucleation of a superstructure under the effect of the Dzyaloshinskii–Moriya interaction.

Formation of the chiral magnetic soliton lattice in metal formates $[NH_4][M(HCOO)_3]$ with $M^{2+}$ = Mn, Fe, Co, Ni, $KCo(HCOO)_3$; and $Cr_{1/3}NbS_2$ occurs under effect of DM forces. According to refs. [51] and [52], since the DM interaction always induces the electric polarization, the true reason of electrical polarization in $[NH_4][M(HCOO)_3]$, $KCo(HCOO)_3$, and $Cr_{1/3}NbS_2$ must be the magnetic one – the formation of chiral magnetic soliton lattice. Activation of DM forces in $[NH_4][M(HCOO)]_3$ must occur at higher temperature than in $KCo(HCOO)_3$ and $Cr_{1/3}NbS_2$, since the ordering of $NH_4^+$ ions and the transition from the $P6_322$ phase to the $P6_3$ one disrupts the centrosymmetric character of the magnetic ions sublattice. Besides, the quasi-one-dimensional character of the metal formates magnetic system is determined by the crystal structure as early as at room temperature, unlike $Cr_{1/3}NbS_2$, in which strength domination of chiral spin helices can be attained only through temperature decrease.

The metastable polymorph $KCo(HCOO)_3$ with the existing, according to our calculations, interrelation between chiral polarization of crystalline and magnetic structures appears to be very interesting from the theoretical point of view. Disruption of the structural chirality (change of the coordination mode of some formates from *anti-anti* to *syn-anti*) in this polymorph is accompanied by disruption of the spin chirality (transformation of one-dimensional chiral helimagnet into three dimensional antiferromagnet). According to ref. [42], this process of transformation of the crystal structure of $K[Co(HCOO)_3]$ from chiral to achiral polymorph is irreversible under normal conditions without the impact of magnetic field. It is important to perform experimental studies in order to understand: Is magnetic field capable to induce the formation of the chiral magnetic soliton lattice in $KCo(HCOO)_3$, which would be accompanied with displacements of atoms transforming the achiral crystal structure into the chiral one? In other words, the reversible transition achiral–chiral polarization of crystalline and magnetic structures in the metastable polymorph $KCo(HCOO)_3$ is possible under effect of magnetic field or not? As a rule, the area of stability of metastable phases is characterized by temperature and pressure [53]. In this case, we consider magnetic phases under effect of not only temperature and pressure, but also that of the external magnetic field. So, we believe it would be reasonable to introduce one more parameter characterizing the magnetic field in the course of examining metastable magnetic phases.

To sum up, we characterized simple in composition, but very interesting objects – metal-formate frameworks of $[NH_4][M(HCOO)_3]$ with $M^{2+}$ = Mn, Fe, Co, and Ni and $KCo(HCOO)_3$. Experimental studies of the dynamics of transformation of chiral helimagnetic structure in magnetic fields and investigation of multiferroics and magnetoelectric effects will enable to evaluate the possibility of using these potential chiral magnetic solitons in spintronics.

**Acknowledgments** The work was partially supported by the Program of Basic Research "Far East" (Far-Eastern Branch, Russian Academy of Sciences), project no. 15-I-3-026.